\documentclass[%
 reprint,
 superscriptaddress,
 amsmath,amssymb,
 aps,
 pra,
floatfix,
]{revtex4-2}
\usepackage[dvipsnames]{xcolor}
\usepackage{graphicx}% Include figure files
\usepackage{dcolumn}% Align table columns on decimal point
\usepackage{bm}% bold math
\usepackage{pgfplots}
\usepackage[T1]{fontenc}
\usepackage{hyperref}% add hypertext capabilities
\usepackage{siunitx}
\usepackage{placeins}
\usepackage[dvipsnames]{xcolor}
\definecolor{mblue}{RGB}{42, 54, 144} % dark blue
\hypersetup{colorlinks=true,breaklinks=true,urlcolor=blue,linkcolor=blue,citecolor=blue}

\begin{document}
\preprint{APS/123-QED}

\title{Site-selective preparation of two-dimensional dipolar quantum gases in an optical beat-note lattice}
% Force line breaks with \\
%\thanks{A footnote to the article title}%

\author{Niclas H\"ollrigl}
\affiliation {Institut für Quantenoptik und Quanteninformation (IQOQI), Österreichische Akademie der Wissenschaften, 6020 Innsbruck, Austria}
\affiliation {Institut für Experimentalphysik, Universität Innsbruck, 6020 Innsbruck, Austria}

\author{Marian Kreyer}
\affiliation {Institut für Quantenoptik und Quanteninformation (IQOQI), Österreichische Akademie der Wissenschaften, 6020 Innsbruck, Austria}

\author{Rudolf Grimm}
\author{Emil Kirilov}
\email{emil.kirilov@uibk.ac.at}
\affiliation {Institut für Quantenoptik und Quanteninformation (IQOQI), Österreichische Akademie der Wissenschaften, 6020 Innsbruck, Austria}
\affiliation {Institut für Experimentalphysik, Universität Innsbruck, 6020 Innsbruck, Austria}

\date{\today}% It is always \today, today,
             %  but any date may be explicitly specified

\begin{abstract}
High-resolution microscopy of two-dimensional dipolar quantum gases requires selecting individual atomic layers, a task complicated for strongly magnetic lanthanide atoms by the limited applicability of standard magnetic-gradient techniques. We present an all-optical method for the deterministic spatial selection of single- and bilayer samples of cold dipolar atoms using spatially selective parametric heating within a beat-note superlattice. By utilizing a high-resolution microscope objective as a common retroreflector for both optical frequency components, the lattice planes are passively stabilized. This renders their positions exceptionally robust against experimental drifts and structure-borne vibrations, even eliminating the need for active laser stabilization over millimeter-scale separations from the reflecting surface. We validate this approach by demonstrating the robust isolation of one or two atomic layers in precise coincidence with the focal plane of our objective. This enables future single-atom-resolved studies of long-range interacting systems.
\end{abstract}

%\keywords{Suggested keywords}%Use showkeys class option if keyword
                              %display desired
\maketitle

\section{Introduction}
A two-dimensional system of particles featuring both short-range contact and long-range dipole-dipole interactions (DDI) exhibits a substantially broader range of many-body phenomena than its conventional, purely contact-interacting counterpart. In the continuum, the addition of DDI enriches the conventional Berezinskii–Kosterlitz–Thouless (BKT) \cite{Berezinskii1971,Kosterlitz1973Apr} picture by mediating anisotropic and long-range vortex–vortex interactions. For large or negative values of the ratio between the dipolar length and $s$-wave contact scattering lengths, $a_\text{dd}/a_s$, they are predicted to host direction-dependent vortex binding and show deviations from the standard contact-dominated BKT behavior~\cite{Mulkerin2013aal}. 

Imposing a periodic potential further enriches the system. Strongly correlated two-dimensional lattices of dipolar particles offer a promising route to simulate extended Hubbard models and itinerant quantum phases governed by long-range anisotropic interactions~\cite{Lagoin2022ebh, Su2023dqs}. In addition to genuinely two-dimensional lattice geometries, anisotropic in-plane lattices can realize arrays of effectively independent one-dimensional tubes by suppressing coupling between neighboring chains. Along each tube, a short-spacing lattice implements extended Hubbard models, where unit-filled systems can host a symmetry-protected topological Haldane insulator and a broader hierarchy of insulating states~\cite{Korbmacher2023lco, Lacki2024gso}. Beyond these fundamental interests, isolating a single two-dimensional layer is an experimental prerequisite, since high-resolution quantum-gas microscopy enables in situ probing in planar geometries~\cite{Gross2021qgm}. Finally, the addition of a second layer gives rise to qualitatively new phenomena, such as interlayer pairing and pair-superfluid phases in lattice geometries~\cite{Safavi-Naini2013qpo}, correlated paired and supersolid states in continuum bilayer gases~\cite{Cinti2017pod} and ultradilute self-bound dipolar liquids~\cite{Guijarro2022uql}. The recent experimental realization of bilayers with magnetic atoms will enable controlled studies of interlayer dipolar coupling~\cite{Du2024apo}.

We report an all-optical method for engineering single- and bilayer two-dimensional atomic samples, combining the species-independent versatility of all-optical manipulation with lattice potentials phase-locked to the front reflective surface of a microscope objective. Our method is based on trapping atoms in a beat-note superlattice (BNSL)~\cite{Masi2021sbo,Petrucciani2025lwo}, within which specific atomic planes are singled out via site-selective parametric heating. By referencing both the trapping potentials and imaging system to the same optical surface, mechanical drifts are largely rejected, ensuring that the sample remains within the submicron depth of focus. We experimentally validate this approach by characterizing the local excitation frequencies of the superlattice and demonstrating the targeted preparation of isolated single- and bi-layer geometries. Additionally, we project a short-period optical lattice onto the two-dimensional atomic sample and measure its site-resolved contrast through the high-resolution objective. The resulting contrast provides a direct and sensitive probe of focal overlap, analogous to an optical modulation transfer function measurement at a single spatial frequency, where even small axial displacements from the focal plane rapidly degrade the observed contrast. Ultimately, as it relies exclusively on external far-detuned optical potentials, our method constitutes a robust tool that applies broadly to diverse cold-atom species, irrespective of their quantum statistics or internal state structure.

This article is structured as follows. In Sec.~\ref{subsec:overview_of_existing_methods} we give an overview of existing methods for the preparation of single atomic layers for quantum-gas microscopy. Section \ref{subsec:beat-note_superlattice_potential} introduces the BNSL potential and discusses regimes relevant to our method. Section \ref{sec:experimental_procedure} then describes the experimental setup, followed by the characterization of the BNSL envelope in Sec.~\ref{subsec:envelope}, and the small spaced lattice in Sec.~\ref{subsec:single_site}. We then demonstrate the preparation of single planes and bilayers in Sec.~\ref{sec:results} and verify that our method prepares single planes coincident with the focal plane of our microscope objective. Finally, we conclude in Sec.~\ref{sec:Conclusions and Outlook} and propose further applications.     

\section{Single-site preparation of quasi-two-dimensional systems}
\subsection{Overview of existing methods}
\label{subsec:overview_of_existing_methods}
A range of experimental techniques has been developed to isolate a single atomic layer. Several of them passively stabilize the trapping potentials to a high-resolution quantum-gas microscope or a superpolished substrate. This referencing provides crucial robustness against structure-borne vibrations and experimental drifts, ensuring the atomic sample remains strictly within the submicron depth of focus of the imaging system. Such methods include combining an evanescent field, magnetic confinement, and the nodal structure of a reflected optical standing wave~\cite{bender2009tsq,Gillen2009tdq}, repulsive optical walls of blue-detuned light, combined with magnetic confinement~\cite{Rath2010eso, Dyke2016cft}, or employing spatially selective microwave transitions within a magnetic field gradient~\cite{Sherson2010sar,Haller2015sai}. The latter technique has further been utilized to implement spin-state-dependent optical removal~\cite{Peaudecerf2019mpo}. 

For strongly magnetic lanthanide atoms, the reliance on large magnetic field gradients renders these techniques broadly unsuitable, as their dense Feshbach resonance spectra make interactions highly sensitive even to small variations of the magnetic field \cite{Chomaz2022}. Furthermore, bosonic isotopes lack the ground-state hyperfine structure required for conventional microwave addressing. Alternatively, all-optical techniques, such as adiabatically compressing an extended cloud using variable-period accordion lattices~\cite{Williams2008dol, Ville2017lac}, circumvent the need for magnetic gradients. Yet, because these potentials are generally not referenced to the microscope objective, they remain highly susceptible to experimental drifts.

%\section{Single-site preparation for quantum-gas microscopy}
%\subsection{Overview of existing methods}

\subsection{Beat-Note Superlattice Potential}
\label{subsec:beat-note_superlattice_potential}
\begin{figure}
\includegraphics[]{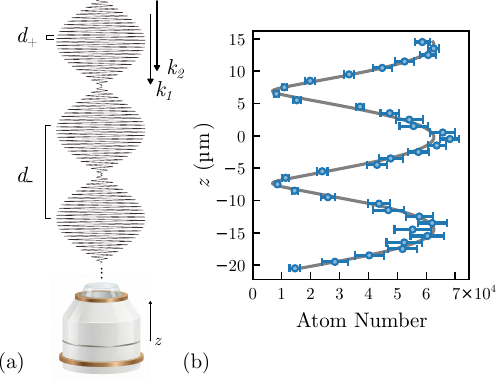}
\caption{\label{fig:loading} Realization of a BNSL. (a) Illustration of the BNSL created by retroreflecting two laser beams with different wavelengths from the top surface of our high NA objective. (b) Measurement of the atom number after loading and holding an atomic sample in the BNSL (blue data points) with a fit based on a heuristic model, proportional to the envelope [see Eq.~\eqref{eq:fit}] (gray line). Error bars represent the error of the mean derived from 10 repetitions of the experiment.}
\end{figure}
Optical dipole forces in standing-wave laser fields with multiple frequency components can show a variety of unusual properties and intriguing phenomena without counterparts in monochromatic fields. The interest in manipulation of atoms with bichromatic standing light waves dates back more than three decades \cite{Kazantsev1987rot, Voitsekhovich1988lpo, Voitsekhovich1989ooa, Kazantsev1989reo, Grimm1990ooa}. Emergent phenomena such as `dipole force rectification’ \cite{Kazantsev1987rot, Kazantsev1989reo, Grimm1990ooa}, the coherent splitting of matter waves \cite{Grimm1994cbs}, or the ‘bichromatic force’ \cite{Voitsekhovich1988lpo, Voitsekhovich1989ooa, Soeding1997sda, Metcalf2017sof} can be interpreted in terms of spatially modulated standing
waves, i.e., beat notes in space, or equivalently as effects of amplitude
modulation, i.e., beat notes in time. Bichromatic optical lattice traps with superperiods in the range of a few 10\,$\mu$m were proposed in \cite{Grimm1995tab} and realized in \cite{Goerlitz2001rob}. 

%Building on these concepts, 
The specific configuration employed in the present work, referred to as a beat-note superlattice (BNSL), is formed by superimposing two collinear optical standing waves with slightly different wavelengths $\lambda_1$ and $\lambda_2$. Applying a similar formalism as in \cite{Masi2021sbo}, one can write the corresponding optical potential as
\begin{equation}
    V_\text{B}(z)=V_0\left[1-\cos(k_- z+\phi_-)\cos(k_+z+\phi_+)\right],
\end{equation}
with $k_{i=1,2} = 2\pi/\lambda_i$, $k_\pm=k_1\pm k_2$ and $\phi_\pm=\phi_1\pm\phi_2$, where $\phi_{i=1,2}$ are the phases of the individual lattices and $V_0=V_1=V_2$ is the depth of the potential, assuming the individual depths $V_{i=1,2}$ are equal. As illustrated in Fig.~\ref{fig:loading}(a), we can identify a closely spaced lattice with lattice spacing $d_+=2\pi/k_+$, which is modulated by a slowly varying envelope of $d_-=2\pi/k_-$.

Much study has already been devoted to the low-energy, perturbative regime where $V_0$ is much smaller than the characteristic energy scale of the short-spaced lattice $E_\text{B+}=\hbar^2k_+^2/8m$. In this work, however, we primarily operate within the tight-binding regime, where $V_0 \gg E_\text{B+}$. In such a setting, tunneling between adjacent sites of the short-spaced lattice is strongly suppressed, resulting in flat bands. Each of these sites can then be approximated by an independent harmonic oscillator whose depth follows the slowly varying envelope
\begin{equation}
    \tilde{V}(z)=V_0\sqrt{\cos^2(k_-z+\phi_-)+\alpha^2\sin^2(k_-z+\phi_-)},
    \label{eq:envelope}
\end{equation}
where $\alpha$ accounts for a possible mismatch $V_{1,2}=V_0(1\pm\alpha)$ due to imperfections in an experimental setup. Thus, in the harmonic oscillator approximation, the local transition frequency $\omega(z)$ becomes equal to the harmonic oscillator transition frequency, and is strongly site dependent, given by
\begin{equation}
    \omega(z) \approx \sqrt{\frac{8E_{\text{B}+}\tilde{V}(z)}{\hbar^2}},
    \label{eq:local_freq}
\end{equation}
for a site located at position $z$. In particular, the frequency difference between neighboring sites can be approximated over a wide range by
\begin{equation}
    \delta(m)=V_0\frac{2\pi^2m}{\hbar(2n+1)^2},
    \label{eq:delta}
\end{equation}
where $m$ is the plane index with the arbitrary choice of $m=0$ situated at the maximum of the BNSL envelope. 

As detailed in the following, we exploit this site-dependent excitation frequency to deterministically remove atoms from unwanted planes via spatially selective parametric heating, allowing for the precise isolation of single- and bilayer two-dimensional quantum gases.
\section{Experimental Setup}
\label{sec:experimental_procedure}
The procedure of loading and cooling the atomic sample to degeneracy follows the same path as described in our previous work~\cite{Anich2024cco}. Bosonic $^{164}$Dy atoms are decelerated in a Zeeman slower and subsequently captured in a compressed magneto-optical trap, operating on the \SI{626}{\nano\meter} transition \cite{Maier2014nlm, Muhlbauer2018soo}. The atoms are then loaded into an optical dipole trap (ODT), formed by two beams crossing at an angle of $\sim\SI{15}{\degree}$. This optical dipole trap is used to transport the atomic sample from the MOT region to a position above the microscope objective. Horizontal transport is achieved by translating the crossing point of the two beams via an aspheric lens mounted on a motorized linear translation stage in their beam path.
For each ODT beam, a pair of cylindrical lenses forms a $2f$ cylindrical telescope, with one of the lenses mounted on a vertical translation stage to enable positioning the sample near the microscope focal plane. During transport we perform evaporative cooling to a BEC, such that, at this stage we typically trap up to \SI{2e5}{} atoms at temperatures of a few tens of \si{\nano\kelvin}. 

The BNSL is formed by two retroreflected laser beams with wavelengths  $\lambda_1 = \SI{1053}{\nano\meter}$ and $\lambda_2 = \SI{1095}{\nano\meter}$, which do not necessarily exactly satisfy the commensurability condition $(n+1)\lambda_1=n\lambda_2$, with $n\approx25$. These specific near-infrared wavelengths were selected based on the availability of laser sources and their compatibility with our microscope objective. The objective features a high-reflectivity coating in this spectral range, allowing the flat front surface of its uppermost lens to act as the retro-reflecting mirror. The two laser fields are spatially overlapped and delivered through a common optical fiber to ensure stable mode matching. A small fraction of the combined beam is sampled before the atoms and diffracted by a holographic grating to separate the wavelengths, allowing independent power stabilization and ensuring a highly reproducible beat-note lattice. Both laser fields are produced by ytterbium-doped fiber amplifiers. The \si{1053}-\si{\nano\meter} amplifier is seeded by a fiber laser, whereas the \si{1095}-\si{\nano\meter} amplifier is seeded by an extended-cavity diode laser (ECDL). To align a single plane of the BNSL with the focal plane of the microscope, situated \SI{160}{\micro\meter} from the retro-reflecting front surface, the ECDL provides the wavelength tunability required to translate the beat-note pattern. 
\begin{figure}
    \centering
    \includegraphics{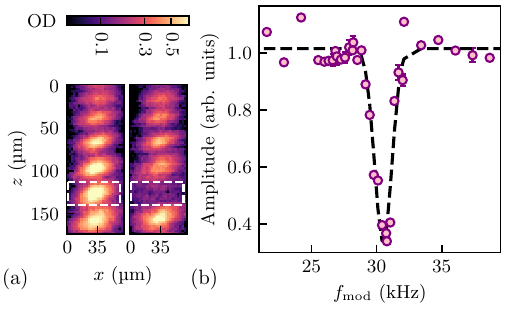}
    \caption{Parametric heating of a single plane in a BNSL. (a)~Absorption images of the atomic cloud after matter-wave magnification ($M \approx 50$), showing resolved planes of the small-spaced lattice before (left) and after (right) site-selective parametric heating of a single plane indicated by the white dashed rectangle. (b)~Site-resolved loss spectrum of the targeted plane. The data points are obtained by fitting the amplitudes of a sum of Gaussian curves to the optical density (OD) integrated along the horizontal direction. Error bars represent the fit uncertainty. The dashed line shows a Gaussian fit to the data.}
    \label{fig:hole}
\end{figure}
\begin{figure}
    \centering
    \includegraphics[width=\columnwidth]{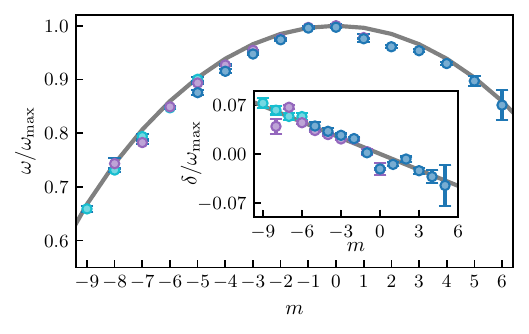}
    \caption{Single-site resolved local transition frequency $\omega$ of the beat-note superlattice, in units of the maximum transition frequency $\omega_\text{max}$, as a function of the lattice plane index $m$. Transition frequencies are extracted from the center of a Gaussian fit to spatially resolved, phase-modulation-induced atom loss spectra. Different colors denote different heights of loading into the BNSL, covering overlapping regions. The solid gray line represents the theoretical prediction based on the harmonic oscillator approximation and is not a fit to the data. Inset: The relative frequency difference between adjacent lattice sites, $\delta/\omega_\text{max}$. A sufficiently large $\delta$ enables the selective preparation of individual planes via parametric heating. This finite difference depends strictly on the wavelength ratio $n$ of the BNSL beams (here $n=25$), with the solid gray line showing the theoretical prediction. Error bars represent the fit uncertainty.}
    \label{fig:frequency_mapping}
\end{figure}
\begin{figure*}
    \centering
    \includegraphics{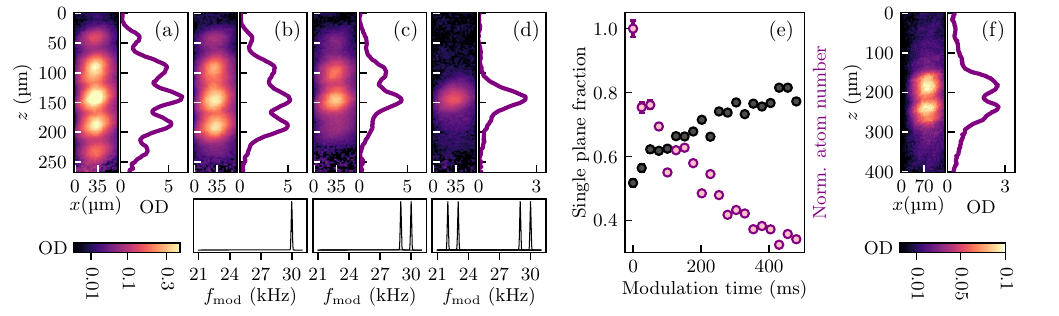}
    \caption{Targeted preparation of isolated single planes and bilayers. (a)-(d) Matter-wave magnified ($M\approx90$) absorption images and integrated optical density (OD) profiles demonstrating the isolation of a single lattice plane. (a) Unperturbed sample after loading. (b)-(d) Progressive removal of atoms from adjacent planes via parametric heating. The specific modulation frequencies $f_{\text{mod}}$ applied to empty the neighboring sites are shown in the corresponding bottom panels. (e) Time evolution of the single-plane preparation, corresponding to the applied frequency spectrum in (d). The fraction of atoms residing in the target plane (black squares, left axis) and the normalized atom number remaining in that plane (purple circles, right axis) are shown as a function of modulation time. Data points are extracted from Gaussian fits to the magnified density profiles. Error bars denote the fit uncertainty. (f) Isolation of a bilayer structure after matter-wave magnification ($M\approx110$). This is achieved by loading atoms near the top of the BNSL envelope, which utilizes a distinct modulation sequence compared to the single-plane preparation.}
    \label{fig:showcase}
\end{figure*}

\section{Characterization of the BNSL}
\subsection{Superlattice envelope}
\label{subsec:envelope}
To characterize the spatial structure of the BNSL, we map its envelope by first loading the atoms from the ODTs into a tightly confining light sheet, whose position along the $z$-axis is controlled via a closed-loop actuator, enabling precise scanning of the atomic sample through the lattice envelope. The sheet is realized at a wavelength of $532\,\mathrm{nm}$ with highly anisotropic waists of $w_x \approx 500\,\mu\mathrm{m}$ and $w_z \approx 5\,\mu\mathrm{m}$, providing strong confinement along the vertical direction with a trap frequency of $\omega_z \approx 2\pi \times 2.5\,\mathrm{kHz}$ at a total power of $\sim\SI{3}{\watt}$.
This configuration localizes the atomic cloud into a thin slice along the $z$-axis while maintaining weak confinement in the transverse plane, enabling spatially resolved probing of the BNSL envelope. For each experimental run, we vary the position of the light sheet along the vertical axis. We then adiabatically ramp down the power of the light sheet while simultaneously increasing the BNSL depth to $12E_\text{B+}$. At this low BNSL power, the vertical potential becomes too shallow near its nodes to support the atoms against gravity, leading to a preferential loss of atoms in these regions. Following a hold time of \SI{200}{\milli\second}, the atoms are released and detected via absorption imaging after a short time of flight. As shown in Fig.~\ref{fig:loading}(b) the resulting atom number $N(z)$ oscillates as a function of the vertical position, showing good agreement with a heuristic model
\begin{equation}
    N(z)=A\sqrt{\cos^2(k_-z+\phi_-)+B},
    \label{eq:fit}
\end{equation}
based on the assumption of a linear dependence of the loaded atom number on the envelope [Eq.~\eqref{eq:envelope}] of the BNSL.  From this fit with four free parameters, we extract the spacing of the BNSL envelope $\pi/k_-=\SI{14.25(7)}{\micro\meter}$ and the position of the nodes, which can be adjusted relative to the focal plane of the objective by adjusting the ECDL wavelength. Close to the focal plane, at a distance of about $\SI{160}{\micro\meter}$ from the retroreflector, the corresponding node of the BNSL shifts with the ECDL wavelength as $\Delta z / \Delta \lambda_2 \approx 3.7\,\mu\mathrm{m}/\mathrm{nm}$. Therefore, no active laser stabilization is required to maintain a stable distance of the individual planes from the microscope objective. The dimensionless parameter $B=\SI{0.012(3)}{}$ incorporates both the optical power mismatch $\alpha$ and the aforementioned gravitational threshold.

\subsection{Site-resolved manipulation}
\label{subsec:single_site}
\begin{figure}
    \centering
    \includegraphics{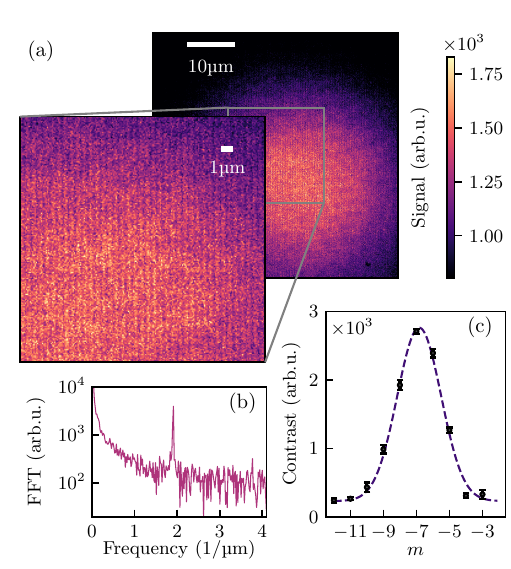}
    \caption{In-situ imaging of a 1D optical lattice. (a) Averaged fluorescence image (10 experimental realizations) of the atomic cloud confined to a single plane of the BNSL, coincident with the focal plane of our objective. Additionally, the atoms are held in a horizontal 1D optical lattice with a lattice spacing of \SI{526.5}{\nano\meter}. (b) One-dimensional Fourier transform of the spatial profile from (a), integrated along the vertical axis. The prominent peak corresponds to the spatial frequency of the horizontal lattice. (c) Lattice contrast as a function of the loaded plane index $m$. The image and FFT on the left correspond to the point at $m = -7$. The dashed line shows a Gaussian fit with constant offset.}
    \label{fig:lattice}
\end{figure}
To resolve the atomic distribution within individual planes of the small-spaced lattice, we need to overcome the optical resolution limit of our side-absorption imaging setup, which cannot directly image the short lattice spacing ($\sim\SI{536.8}{\nano\meter}$). We achieve this by employing an all-optical matter-wave magnifying protocol \cite{Tung2010oot, Murthy2014mwf,Asteria2021qgm,Brandstetter2025mtw} to separate the individual samples, as shown in Fig.~\ref{fig:hole}(a). The matter-wave lensing technique exploits the phase-space rotation generated by harmonic confinement. After a quarter-period evolution in a harmonic trap,
\( t=\pi/(2\omega_1) \), the initial real space distribution is mapped onto momentum space. For this we use the same anisotropic light sheet trap used
earlier for mapping the BNSL envelope, with $\omega_1 = 2\pi\times\SI{2.5}{\kilo\hertz}$. We then switch off all traps and let the atoms expand over a time of flight $t_\text{TOF}$, yielding an effective magnification proportional to \( \omega_1 t_{\mathrm{TOF}} \). We choose $t_\text{TOF}$ such that the expanded sample fits in our imaging region, with typical values for the total magnification factor $M$ ranging from 50 to 110.

%A subsequent quarter-period evolution in a second trap with frequency \( \omega_2 \) performs the inverse transformation, mapping the distribution back into coordinate space with a spatial magnification given by the ratio \( \omega_1/\omega_2 \). 
%In the regime relevant to our work, the second harmonic evolution can instead be replaced by free expansion over a time of flight \( t_{\mathrm{TOF}} \), yielding an effective magnification proportional to \( \omega_1 t_{\mathrm{TOF}} \). The harmonic confinement is provided by the same anisotropic light sheet trap used earlier for mapping the BNSL envelope, resulting in magnification exclusively along the tightly confined vertical direction. 

For a fixed atom number, the BNSL populates fewer lattice planes than a single-wavelength lattice, since the beat-note envelope creates an additional axial potential such that only regions where the local chemical potential $\mu - \tilde{V}(z)$ lies above the local band minimum are occupied. Depending on the atom number and the position along the BNSL envelope, we typically load 5--10 planes by linearly extinguishing the ODTs while simultaneously ramping up the BNSL
over a duration of \SI{100}{\milli\second}. At the antinode of the BNSL,
the short-period lattice reaches a depth of approximately $320\,E_\text{B+}$. 

We then probe the local transition frequency by modulating the amplitude of one of the BNSL beams. In a beat-note lattice formed by two standing waves with different wavevectors, amplitude modulation of a single beam shifts the relative phase $\phi_-$ of the two optical potentials and periodically
displaces the lattice minima by an amount scaling as
$\delta z \propto (k_1-k_2)/(k_1k_2)$. This introduces a linear
perturbation $\propto z$ of the local harmonic confinement that
couples the $0\!\rightarrow\!1$ vibrational states, while amplitude modulation in a single-frequency lattice would predominantly drive the $0\!\rightarrow\!2$ parametric resonance. We tune the modulation frequency to the $0\!\rightarrow\!1$ transition and later remove the excited atoms from the trap via a brief evaporation stage, achieved by lowering the power of both BNSL beams.

By tuning the modulation frequency, we can selectively address and remove atoms from specific planes. Extracting the remaining atom number from spatial regions of interest yields site-resolved loss spectra, as shown in Fig.~\ref{fig:hole}(b). The local transition frequency is then determined from the center frequency of a Gaussian fit to these spectra, allowing us to assign a discrete plane index $m$, as shown in Fig.~\ref{fig:frequency_mapping}. As expected, the measured spatial variation of the local transition frequency is in agreement with Eq.~\eqref{eq:local_freq}. As can also be seen from Eq.~(\ref{eq:delta}), the difference between neighboring planes is smallest at the maximum of the envelope, but loading at the side, where the energy differences are larger, allows for a better discrimination between planes. 

\section{Preparation of single planes and bilayers}
\label{sec:results}
We now utilize the site-resolved parametric heating characterized in the previous section to physically engineer the density distribution of the sample. By applying a sequence of amplitude modulation pulses to the unperturbed sample shown in Fig.~\ref{fig:showcase}(a), we can systematically target and empty unwanted lattice planes one by one. Fig.~\ref{fig:showcase}(b) shows the removal of the lowest plane by modulating the lattice with one frequency. By adding additional modulation steps with different frequencies [(c) and (d)] we are left with most of the atoms in a single plane. Fig.~\ref{fig:showcase}(f) shows how the normalized atom number and fraction of atoms in the center slice depend on the time modulation is applied. The control over the loading position also allows us to prepare a bi-layer system, see Fig.~\ref{fig:showcase}(e). For this, we load at the maximum of the envelope, where neighboring planes are almost degenerate in energy. Single-plane preparation is achieved best by loading the atoms at the slope of the BNSL and performing a tailored sequence of parametric-heating pulses from above and below the target plane.

To verify that an isolated single plane resides within the depth of focus of the high-resolution objective, we characterize the system performance using in-situ fluorescence imaging. The atoms are illuminated by a train
of alternating counterpropagating pulses, applied one at
a time, with a total duration of $1\,\mu\mathrm{s}$ to limit
diffusion due to photon scattering \cite{Anich2024cco}. We systematically prepare the atomic sample in different vertical planes, enumerated by the index $m$, and additionally confine the atoms within a horizontal 1D optical lattice with a spacing of \SI{526.5}{\nano\meter}. As illustrated in Fig.~\ref{fig:lattice}(a) for the optimal plane, this reveals a clear spatial modulation of the atomic density. To quantify this, we integrate the spatial profile along the vertical axis and calculate the one-dimensional Fourier transform, as shown in Fig.~\ref{fig:lattice}(b). We then extract the imaging contrast from the height of the prominent spatial frequency peak corresponding to the horizontal lattice spacing. As shown in Fig.~\ref{fig:lattice}(c), the extracted lattice contrast exhibits a sharp profile with \( \sigma = 0.70(3)\,\mu\mathrm{m} \) around the optimally aligned plane ($m= -7$), rapidly decaying for adjacent planes as they move out of the depth of focus, confirming our ability to deterministically position the 2D sample for high-resolution quantum-gas microscopy \cite{sohmen2023asi}.

We perform time-of-flight thermometry immediately after loading the BNSL and after selecting a single plane by parametric heating. For the former we utilize again our vertical matter-lensing magnification (as in Fig.\ref{fig:showcase}(a)), where after a time of flight of \SI{15}{ms} we analyze the lateral orthogonal to the lensing direction cloud profile, which is well described by a bimodal distribution. The thermal Gaussian component yields \SI{123(2)}{nK}. After the parametric heating the temperature in a selected plane is slightly elevated and we observe density fluctuations caused by thermal 2D phase fluctuations \cite{Dettmer2001oop}. At this point we reconstruct the 2D temperature by extracting the in-plane momentum distribution of the selected two-dimensional layer using this time in plane matter-wave lensing.  
In quasi-two-dimensional geometry, the dimensionless effective interaction strength is given by 
$\tilde g_{\mathrm{eff}}=\sqrt{8\pi}\,\bigl(a_s + a_{dd}(3\cos^2\theta - 1)\bigr)/l_z \approx 1.9$, 
where for our $^{164}$Dy system the $s$-wave scattering length is $a_s = (100 \pm 5)\,a_0$, the dipolar length is 
$a_{dd}=(\mu_0 \mu^2 m)/(12\pi \hbar^2)\approx130\,a_0$, $\theta=0$ denotes the angle between the dipole orientation and the normal to the two-dimensional plane, and $\omega_z/2\pi=25\,\mathrm{kHz}$ is the axial confinement frequency defining $l_z=\sqrt{\hbar/(m\omega_z)}$. We perform both Boltzmann and polylogarithmic ideal quantum gas fits in the region of the momentum distribution $(p^2)/(2 m k_B T) \gtrsim \tilde{g}$. The temperature derived by this method is \SI{180\pm 5}{nK}. For typical atom numbers of $N=10^4$, the system is deeply in the quasi-two-dimensional regime, since $N \tilde g_{\mathrm{eff}}/\sqrt{8\pi} \, (\omega_r/\omega_z)^2 \ll 1$ and $k_B T \ll \hbar\omega_z$, ensuring frozen dynamics along the tightly confined direction.

\section{Conclusion and Outlook}
\label{sec:Conclusions and Outlook}
In conclusion, we have developed a method to isolate a two-dimensional gas of dipolar particles based on parametric heating in a BNSL. The selected two-dimensional potential corresponds to a single, well-defined BNSL plane, which is intrinsically referenced to the front surface of a microscope objective, thereby suppressing technical drifts and mechanical vibrations. Once the two-dimensional potential is tuned to the focal plane of the microscope, no active phase stabilization of the BNSL lasers is required, even for retroreflector distances on the millimeter scale. The layer isolation protocol relies exclusively on optical dipole potentials, thereby avoiding the need for magnetic field gradients or microwave-assisted cleaning of lattice planes. As a result, the technique is applicable to various cold-atom species, as it depends only on the optical polarizability at the BNSL wavelengths and is independent of the quantum statistics of the particles or their internal state structure. 

Our method enables precise positioning of atoms near surfaces, providing access to measurements of Casimir--Polder forces that require extremely accurate control of the atom--surface distance \cite{Sukenik1993mot, Bender2010dmo}. The method also naturally enables the realization of bilayer systems. By choosing shorter lattice wavelengths and a sufficiently large wavelength difference between the BNSL beams (small $n$), one can obtain small inter-layer spacing at the \SI{100}{\nano\meter} level and therefore substantial interlayer DDI even for magnetic atoms such as dysprosium.
Finally, the same approach can be used to generate highly isolated one-dimensional tubes of atoms by adding a horizontal BNSL coincident with the focal plane of the microscope. When combined with in-plane matter-wave magnification techniques along the tubes, this system provides an attractive platform for the investigation of exotic one-dimensional phenomena.

\section*{Acknowledgments}
\label{sec:Acknowledgments}
We thank M. Fattori for sharing his expertise and for helpful discussions regarding the BNSL.
This research was funded in part by the Austrian Science Fund (FWF) [10.55776/COE1], the Julian Schwinger Foundation [JSF-22-05-0010] and NextGeneration EU Grant AQuSIM through the Austrian Research Promotion Agency (FFG) (No. FO999896041).

\section*{Data Availability}
The data presented in this article are openly available~\cite{data}. Additional raw data is available from the authors upon reasonable request.
\bibliography{ultracold}

\end{document}